\documentclass[twocolumn,showpacs,aps,prl,floatfix,%
superscriptaddress]{revtex4}  

\usepackage{graphicx} 

\newcommand{\ee}{\end{equation}} 
\newcommand{\eq}{{\,=\,}} 
\newcommand{\bm}{\bf}
\newcommand{\el} 
{\left .\frac{dE}{dx}\right|_0}
 
\begin{document} 
 
 
\title{Jet quenching and elliptic flow }
  
\date{\today}
  
\author{A. K. Chaudhuri} 
\email{akc@veccal.ernet.in} 
\affiliation{Variable Energy Cyclotron Centre, 1-AF, Bidhan Nagar,
Kolkata - 700 064, India} 
 
\begin{abstract} 

In jet quenching, a hard QCD parton, before fragmenting into a jet of hadrons, deposits a fraction of its energy in the medium, leading to  suppressed production of high-$p_T$ hadrons. 
Assuming that the deposited  energy quickly thermalizes, we simulate the subsequent hydrodynamic 
evolution of the QGP fluid. Explicit simulation of Au+Au collision
with and without a quenching jet indicate that elliptic flow is greatly reduced in a jet event. The result can be used to identify the
jet events in heavy ion collisions.
 \end{abstract} 
 
\pacs{PACS numbers: 25.75.-q, 13.85.Hd, 13.87.-a} 
 
\maketitle 
 

The three most important results  that came out from the heavy ion
programme at RHIC are  (i) dramatic suppression of inclusive hadrons production at large transverse momentum (high $p_T$ suppression) \cite{Adcox:2001jp,Adler:2002xw,Adams:2003kv},
(ii) disappearance of away side two hadron correlation peak
\cite{STARjetqu} and   (iii) large elliptic flow \cite{elliptic}.  
High $p_T$ suppression confirmed the theoretical prediction of jet quenching \cite{QGP3jetqu} . 
Long before the RHIC Au+Au collisions, it was predicted, in a pQCD calculation, that in a deconfined medium, high-speed partons will suffer energy loss, leading to suppressed production of hadrons. The observed high $p_T$ suppression in Au+Au collisions \cite{Adcox:2001jp,Adler:2002xw,Adams:2003kv} are in agreement with the prediction.  Large elliptic flow observed in non-central Au+Au collisions confirms fluid like behavior of the matter produced . 
Elliptic flow measured the momentum anisotropy of the produced
particles. In non-central collisions between two identical spherical
nuclei, the reaction zone is spatially asymmetric.  
Rescattering process among the produced particles 
(locally isotropic in momentum space) transfers this spatial asymmetry into the momentum space and momentum distribution of produced particles become anisotropic.  Naturally, in a central
collision between identical spherical nuclei, e.g. Au+Au, reaction
zone is azimuthally symmetric and elliptic flow vanishes.
A large variety of RHIC data are well explained in an
ideal hydrodynamic model, with initial energy density of deconfined matter $\varepsilon_i \sim$ 30 $GeV/fm^{3}$, thermalized at an initial time $\tau_i$=0.6 fm \cite{QGP3v2} .
All these observations are being treated 
as evidences for the creation of a very dense, color opaque medium of deconfined quarks and gluons \cite{QGP3jetqu}.

If the partons lose energy in the medium, what happened to that
energy? It has been suggested that a fraction of lost energy will go to collective excitation, call the
"conical flow" \cite{Stoecker:2004qu,shuryak}. The parton moves with speed of light, much greater
than the speed of sound of the medium ($c_{jet} >> c_s$), and the quenching jet can produce a shock wave with Mach cone angle, $\theta_M=cos^{-1}c_s/c_{jet}$. Resulting conical flow will have characteristic peaks at $\phi=\pi-\theta_M$ and $\phi=\pi + \theta_M$. Indication of such peaks are seen in azimuthal distribution of secondaries associated with high $p_T$ trigger in central Au+Au collisions \cite{Wang:2004kf,Jacak:2005af}. As Mach cone is sensitive to the speed of sound of the medium, it raises the possibility of measuring the speed of sound of the deconfined matter of Quark-Gluon-Plasma. However, theoretical calculation \cite{Satarov:2005mv}, as well as 
explicit simulation of hydrodynamic
evolution of QGP fluid with a quenching jet \cite{Chaudhuri:2005vc,Chaudhuri:2006qk}, 
 indicate that unlike
in a static medium, in a moving fluid, Mach shock fronts are
distorted.  Interplay of fluid velocity, shock velocity, together with the inhomogeneity of the medium, render the simple equation for Mach cone angle, $\theta_M=cos^{-1}c_s/c_{jet}$ invalid.
Peaks seen in the azimuthal distribution of secondaries
in STAR and PHENIX experiments \cite{Wang:2004kf,Jacak:2005af}, even though are due to shock wave production, they may not directly lead to the speed of sound of the QGP fluid.

Other than producing a "conical flow", partonic energy loss in the medium will also affect the sensitive observable, the elliptic flow. Most sensitive manifestation could be  negative elliptic flow  in a  central (zero impact parameter) collision. The reason is simple.
Without a quenching jet, in a central collision, reaction zone is symmetric  and elliptic flow is identically zero. With a quenching jet, azimuthal symmetry is lost
(quenching jet defines a direction)
and elliptic flow can develop. Explicit simulation of hydrodynamic evolution of fluid, with a quenching jet, acting as
a source of energy-momentum, indicate that  
azimuthally symmetric initial energy density (or equivalently temperature)
distribution gets severely distorted and even in a zero impact parameter collision, produced particles show  dependence on the azimuthal angle \cite{Chaudhuri:2005vc,Chaudhuri:2006qk}.
Indeed, if the quenching jet induce shock wave propagation,
one can expect negative flow. For example,
in a medium of hadronic resonance gas (the squared speed of sound $\sim$0.15) 
the Mach angle is $\theta_M=arccos(c_s/c_{jet}) \sim 67^\circ$. The shock wave will inhibit particle production in the angular region $-67^\circ$ to $+67^\circ$ \cite{Chaudhuri:2005vc}. Elliptic flow, which is spectra weighted average of azimuthal angle,
$v_2=<cos(2\phi)>$, will be negative. Negative elliptic flow
in zero impact parameter collision can be an indirect proof of the shock wave production.
 
Let us  consider energetic of jet quenching and elliptic flow. As discussed in \cite{shuryak},  total transverse energy of all the secondaries per one unit of rapidity, in RHIC Au+Au collisions, is, $\frac{dE_T}{d\eta} \sim$ 600 GeV. Most of it is thermal, only a fraction of the transverse energy $\sim$ 100 GeV, goes to collective excitations. Elliptic flow measured in Au+Au collisions $v_2 \sim $0.1. Then of the total transverse energy, only $\sim$ 10 GeV goes to anisotropic excitations. 
Detailed simulations \cite{Chaudhuri:2005vc} indicate that, 
 in central Au+Au collisions  at RHIC energy, a quenching jet, with energy loss appropriate  for observed high $p_T$ suppression, deposits $\sim$ 10 GeV energy  to the system. Energy deposited by the quenching jet   compare favorably with the anisotropic excitation energy.  
The change induced by a quenching jet on elliptic flow could be measured experimentally.

\begin{figure}[h] 
\includegraphics[bb=75 276 531 759
,width=0.80\linewidth,clip]{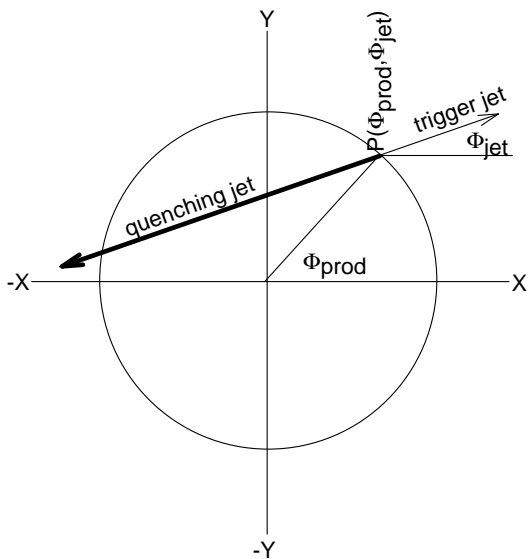}
\caption{Schematic representation of a jet moving through the
medium. The high $p_T$ pair is assumed to produce on the surface of the fireball characterized by the angle
$\phi_{prod}$. One of the jet escapes forming the trigger jet, the
other move in the fireball at an angle $\phi_{jet}$.}
\label{F1}
\end{figure}

Elliptic flow developed due to a quenching jet will depend
on the energy deposited by the jet i.e. on the jet path length.
A schematic representation
of the jet moving through the medium is shown in Fig.\ref{F1}.
We assume that just before hydrodynamics become applicable, a pair of high-$p_T$ partons is produced. At RHIC, hydrodynamics
become applicable quite early, $\tau_i\sim 0.6 fm$ \cite{QGP3v2} .
Time $\tau_i\sim 0.6 fm$ is elapsed between the early
hard collisions producing high-momentum partons and thermalisation. Can the partons lose significant amount of energy in between this time? In \cite{Wang:2004dn} parton energy loss
was estimated. On
the average, energy lose for a 10 GeV parton, in an expanding medium is $<dE/dL>_{1d}\approx 0.85\pm 0.24$ GeV/fm. Then
in the pre-equilibrium stage, parton energy loss is only $\sim  0.51 \pm 0.14$ GeV only \cite{Wang:2004dn}, small fraction of total energy ($\sim$10 GeV) loss. Indeed, it could be even less.  
Strong jet quenching and survival of the trigger jet, forbids production
in the interior of the fireball. Jet pairs can be produced only on a thin shell on the surface of the fireball, where medium density is
much less than the average density. For Au+Au collisions at
impact parameter $b$, we assume that the di-jet is produced on  
the surface of the ellipsoid with minor and major axis, $A=R-b/2$ and
$B=R\sqrt{1-b^2/4R^2}$, with $R=6.4$fm.
 Then jet production point can be characterized by the angle $\phi_{prod}$ 
($-\pi \leq \phi_{prod} \leq +\pi$) only .
One of the jet moves outward  and escapes, forming the trigger jet. The other enters into the fireball. Presently, we assume that the
jet has enough energy to pass through the model.  The trajectory of the jet can be designated by the angle $\phi_{jet}$,
($-\frac{\pi}{2} \leq \phi_{jet} \leq +\frac{\pi}{2}$).
The fireball is expanding and cooling. The ingoing parton travels 
at the speed of light and loses energy in the fireball which 
thermalizes and acts as a source of energy and momentum for 
the fireball medium. We solve the energy-momentum conservation equation,

\begin{equation} \label{1}
\partial_\mu T^{\mu\nu}=J^\nu,
\end{equation}

\noindent where the source is modeled as,

\begin{eqnarray} 
\label{2} 
 &&J^\nu(x)=J(x)\,\bigl(1,-cos(\phi_{jet}),-sin(\phi_{jet}),0\bigr),\\
\label{3}
 &&J(x) = \frac{dE}{dx}(x)\, \left|\frac{dx_{\rm jet}}{dt}\right| 
          \delta^3(\bm{r}-\bm{r}_{\rm jet}(t)).
\end{eqnarray} 
Massless partons have light-like 4-momentum, so the current $J^\nu$
describing the 4-momentum lost and deposited in the medium by the 
fast parton is taken to be light-like, too. $\bm{r}_{\rm jet}(t)$ is 
the trajectory of the jet moving with speed $|dx_{\rm jet}/dt|\eq{c}$.
$\frac{dE}{dx}(x)$ is the energy loss rate of the parton as it moves 
through the liquid. It depends on the fluid's local rest 
frame particle density. Taking guidance from the 
phenomenological analysis of parton energy loss observed in Au+Au 
collisions at RHIC \cite{Eloss} we take
\begin{equation}
\label{4}
  \frac{dE}{dx} = \frac{s(x)}{s_0} \left.\frac{dE}{dx}\right|_0
\end{equation}
where $s(x)$ is the local entropy density without the jet.  
The measured suppression of high-$p_T$ particle production in Au+Au 
collisions at RHIC was shown to be consistent with a parton energy 
loss of $\left.\frac{dE}{dx}\right|_0\eq14$\,GeV/fm at a reference 
entropy density of $s_0\eq140$\,fm$^{-3}$ \cite{Eloss}. 
For  $\left.\frac{dE}{dx}\right|_0\eq14$\,GeV/fm, the jet  deposits $\sim$ 10 GeV to the fireball as it pass through the fireball \cite{Chaudhuri:2005vc}. Most of the energy deposited is early in the evolution. The energy loss is weighted
by the entropy density Eq.\ref{4}. Late in the evolution entropy density decreases to small values and  energy transfer is inefficient. 
It maybe mentioned that even though in Eq.\ref{4} partonic energy loss is assumed to depend only on the fluid rest frame density, it could as well depend on velocity of the parton. In BDMPS \cite{Baier:2000mf}, the average partonic
energy loss could be calculated as,

\begin{equation}
\Delta E= \alpha_s \frac{1}{2} \hat{q} L^2
\end{equation}

\noindent where $\hat{q}=\mu^2/\lambda$ is a transport coefficient
dependent on the medium property (mean free path $\lambda$,
screening length $\mu$).

For the hydrodynamic evolution we use a modified version of the publicly available hydrodynamic code AZHYDRO \cite{QGP3v2,AZHYDRO}. The code
  is formulated in 
$(\tau,x,y,\eta)$ coordinates, where $\tau{=}\sqrt{t^2{-}z^2}$ is the
longitudinal proper time, 
$\eta{=}\frac{1}{2}\ln\left[\frac{t{+}z}{t{-}z}\right]$ 
is space-time
rapidity, and $\bm{r}_\perp{\,=\,}(x,y)$ defines the plane transverse to the 
beam direction $z$. AZHYDRO employs longitudinal boost invariance 
along $z$ but this is violated by the source term (\ref{3}). We 
therefore modify the latter by replacing the $\delta$-function 
in (\ref{3}) by

\begin{eqnarray}
\label{5}
  \delta^3(\bm{r}-\bm{r}_{\rm jet}(t)) &\longrightarrow&
  \frac{1}{\tau}\,\delta(x-x_{\rm jet}(\tau))\,\delta(y-y_{\rm jet}(\tau))
\nonumber\\
 &\longrightarrow&\frac{1}{\tau} \, 
\frac{e^{-(\bm{r}_\perp-\bm{r}_{\perp,{\rm jet}}(\tau))^2/(2\sigma^2)}}
      {2\pi\sigma^2}
\end{eqnarray}
with $\sigma{\,=\,}0.70$\,fm. 
Dependence on the Gaussian width $\sigma$
was studied in \cite{Chaudhuri:2005vc}. With the quenching jet, constant energy density contours show Mach cone like structure, the Mach cone angles get  better defined if the width is reduced
by half. The azimuthal distribution of $\pi^-$ on the other hand remain nearly unaltered. 
Intuitively, the replacement of the 'delta' function by a Gaussian
 replaces the ``needle'' 
(jet) pushing through the medium at one point by a ``knife'' cutting the
medium along its entire length along the beam direction. 
Thus assumption of boost-invariance will over estimate the
effect of jet quenching. Boost-invariance implicitly assume that
the deposited energy is integrated along the rapidity axis.
PHOBOS experiment \cite{Back:2002wb} indicate that in 0-6\% centrality collisions, over a rapidity range $\sim$ (-6 to 6), boost-invariance is approximately valid in the rapidity range $\sim$ (-2.5 to +2.5). The assumption of boost-invariance then overestimate the particle yield by a factor $\sim$1.5. Effect of jet quenching will be overestimated by a similar factor.  In less central collisions, boost-invariance is valid over extended rapidity region and the effect will be still less overestimated.
 
The modified hydrodynamic equations in $(\tau,x,y,\eta)$ coordinates
read \cite{Chaudhuri:2005vc,AZHYDRO}
%
\begin{eqnarray} 
\label{6} 
  \partial_\tau \tilde{T}^{\tau \tau} + 
  \partial_x(\tilde{v}_x \tilde{T}^{\tau \tau}) +
  \partial_y(\tilde{v}_y \tilde{T}^{\tau \tau}) 
  &=& - p + \tilde{J},
\\ 
\label{7} 
  \partial_\tau \tilde{T}^{\tau x} +
  \partial_x(v_x \tilde{T}^{\tau x}) +
  \partial_y(v_y \tilde{T}^{\tau x}) 
  &=& - \partial_x \tilde{p} - \tilde{J}, \quad
\\ 
\label{8} 
  \partial_\tau \tilde{T}^{\tau y} +
  \partial_x(v_x \tilde{T}^{\tau y}) +
  \partial_y(v_y \tilde{T}^{\tau y}) 
  &=& -\partial_y \tilde{p},  \quad
\end{eqnarray}  
%
where $\tilde{T}^{\mu\nu}\eq\tau T^{\mu\nu}$, 
$\tilde{v}_i{\eq}T^{\tau i}/T^{\tau\tau}$,
$\tilde p\eq\tau p$, and $\tilde{J}\eq\tau J$.

To simulate central Au+Au collisions at RHIC, we use the standard 
initialization described in \cite{QGP3v2} and provided in the 
downloaded AZHYDRO input file \cite{AZHYDRO}, corresponding to a 
peak initial energy density of $\varepsilon_0 \eq 30$ $GeV/fm^3$ at
$\tau_0 \eq 0.6$ $fm/c$. We use the equation of state EOS-Q 
 described 
in \cite{QGP3v2,AZHYDRO} incorporating a first order phase transition 
and hadronic chemical freeze-out at a critical temperature 
$T_c{\,=\,}164$\,MeV. The hadronic sector of EOS-Q is soft with a 
squared speed of sound $c_s^2 \approx 0.15$. 

Let us first investigate the evolution of spatial eccentricity and momentum anisotropy in a central ($b=0)$ Au+Au collision, with a quenching jet moving along the x-axis ($\phi_{prod}=\phi_{jet}=0$).   Spatial eccentricity is defined as,

\begin{equation}
\varepsilon_x (\tau)= \frac{<y^2 - x^2>}{<y^2 + x^2>} \label{eq8}
\end{equation}

\begin{figure}[h] 
\includegraphics[bb=23 289 527 770
,width=0.80\linewidth,clip]{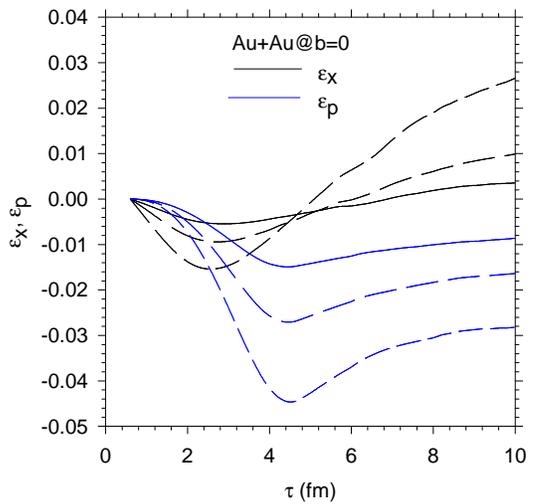}
\caption{ (color online) Evolution of spatial eccentricity (black lines) and momentum anisotropy (blue lines) with time, in a central Au+Au collision . The solid, dashed and short dashed lines corresponds to jet energy loss parameter $\el$=14, 28 and 56 GeV/fm  respectively.}
\label{F2}
\end{figure}

\noindent where  the average in Eq.\ref{eq8} is taken over
the energy densities. 
In Fig.\ref{F2}, 
black lines show the temporal evolution of spatial eccentricity ($\varepsilon_x$) for  three different values for the energy loss parameter $\el$=14, 28 and 56 GeV/fm. 
In a central collision, initially the reaction zone is symmetric and
$\varepsilon_x$=0.  
Without any quenching jet, $\varepsilon_x$
remain zero during the evolution. But as shown in Fig.\ref{F2},
with a quenching jet, $\varepsilon_x$ quickly becomes negative, reaching a maximum negative value around $\tau \sim$2.5 fm. Then the
eccentricity increases, becomes positive around $\tau \sim$5 fm, and continues to increase. If the jet lose more energy, eccentricity
becomes more negative initially and more positive at later time.
The temporal behavior of $\varepsilon_x$ can be understood.
Initially 
the jet is at $(x_{jet},y_{jet})$=(6.4fm,0 fm). Due to energy deposition by the quenching jet,
in the beginning of the evolution $x^2$ term in Eq.\ref{eq8} gets more weight
and $\varepsilon_x$ is negative. But later, as seen in 
\cite{Chaudhuri:2005vc,Chaudhuri:2006qk}, 
constant energy density contours are pushed inside and $x^2$ term gets less weight and $\varepsilon_x$ become negative.

In Fig.\ref{F2}, the blue lines show the temporal evolution of the momentum anisotropy. Momentum anisotropy is  defined as,

\begin{equation}
 \varepsilon_p(\tau)=  \frac{\int dx dy (T^{xx} - T^{yy})}
{\int dx dy (T^{xx} + T^{yy})}
\end{equation}

Like the spatial eccentricity, the momentum anisotropy quickly become negative, but unlike $\varepsilon_x$,  it remain negative at later time also.   Thus at late time  there is a net momentum flow in the y-z plane. Evolution of   momentum anisotropy  clearly indicate that  in a central Au+Au collision at RHIC energy, a
quenching jet can induce negative  elliptic flow.

\begin{figure}[h] 
\includegraphics[bb=25 261 527 770
,width=0.80\linewidth,clip]{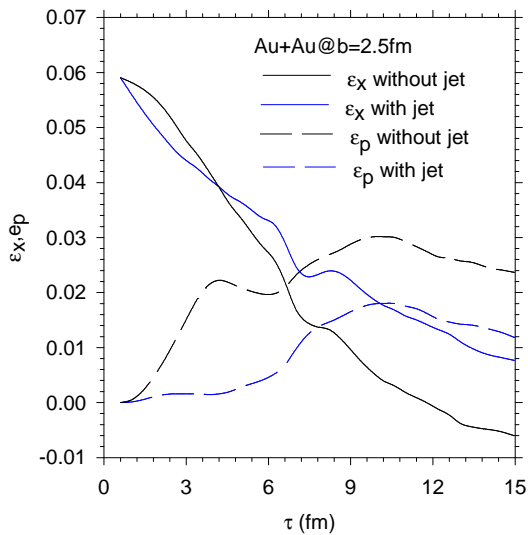}
\caption{ Evolution of spatial eccentricity (solid lines) and momentum anisotropy (dashed lines) with time, in a  Au+Au collision at impact parameter b=2.5 fm. The black and blue lines are for evolution without a jet and with a jet respectively. The jet energy loss parameter $\el$=14  GeV/fm.}
\label{F3}
\end{figure}

However, an experiment cannot be confined to $b=0$ collisions only.
Impact parameter fluctuations have to be allowed for. Even if a 
quenching jet induce negative flow in a b=0 collision, it cannot be measured experimentally. In a finite impact parameter collision,
reaction zone is asymmetric and one obtain positive elliptic flow,
more asymmetric the reaction zone, more is the elliptic flow.
With a quenching jet we expect reduction in elliptic flow.
We now test this expectation by explicit simulation of Au+Au collision at impact parameter b=2.5 fm.
It roughly corresponds to 0-5\% centrality collision. For comparison, we also simulate a b=2.5 fm Au+Au collisions without any quenching jet.
In Fig.\ref{F3}, we have shown the temporal evolution of
the spatial eccentricity (solid line) and momentum anisotropy
(the dashed lines). The blue and black lines corresponds to evolution with and 
without any quenching jet respectively.
  For the jet energy loss we have used $\el$=14 GeV/fm, the phenomenologically acceptable value consistent with
high $p_T$ suppression.  Without any jet, in b=2.5 fm Au+Au 
collision, as the fluid evolves, initially non-zero spatial eccentricity decreases and momentum anisotropy grow. 
With a quenching jet, during the first 4-5 fm, spatial eccentricity decrease more quickly and   the momentum anisotropy grows less with time. The results
corroborate our speculation that with a quenching jet, elliptic flow will decrease.
 
\begin{figure}[h] 
\includegraphics[bb=32 265 524 769
,width=0.80\linewidth,clip]{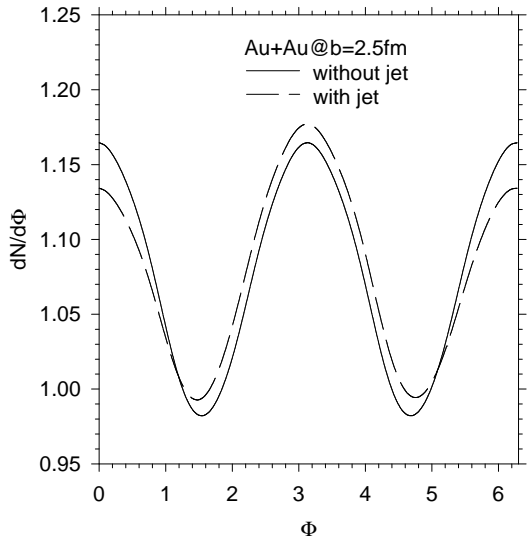}
\caption{ Azimuthal dependence of $p_T$ integrated ($1 GeV\leq p_T \leq 2.5$) pion. The solid and the dashed lines correspond to evolution without a jet and with a jet respectively.
 The jet energy loss parameter $\el$=14  GeV/fm.}
\label{F4}
\end{figure}

Using the standard Cooper-Frey prescription,
for each jet trajectories ($\phi_{prod}$,$\phi_{jet}$)  
 we now calculate the pion spectra $\frac{dN}{d^2p_T d\phi}$, 
at freeze-out temperature of 100 MeV. 
For the jet energy loss, we have used
the phenomenologically acceptable value, $\el$=14 GeV/fm. 
Finally we average over the
jet production angle $\phi_{prod}$ and the jet angle $\phi_{jet}$.  
In Fig.\ref{F4}, azimuthal dependence of $p_T$ ($1 GeV\leq p_T \leq 2.5 GeV$) integrated pion spectra is shown. The solid line is obtained when there is no quenching jet. The dashed line corresponds to evolution with a quenching jet. With a quenching jet, pion production is enhanced at $\phi=\pi$. We also note that pion production is
depleted at $\phi=0$. 
It suggests that in evolution of the fluid with a quenching jet, azimuthal distribution of pions contain a $-cos\phi$ component,
which is absent in evolution without a quenching jet.
It appears that with a quenching jet depositing energy to the medium, directed flow develops.  We will discuss this issue later.

\begin{figure}[h] 
\includegraphics[bb=12 291 559 769
,width=0.80\linewidth,clip]{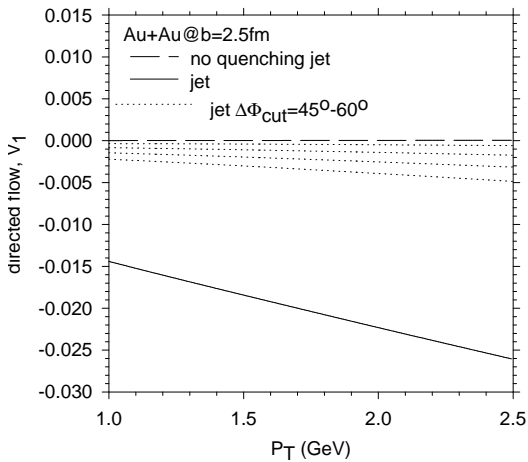}
\caption{Transverse momentum dependence of the directed
flow ($v_1$) in a b=2.5 fm Au+Au collision. Due to symmetry, directed flow ($v_1$) is zero  without any quenching jet (the
dashed  line).
The  solid line shows  the directed flow in presence of a quenching jet. The  dotted lines are directed flow after correcting for the neglect of trigger jet fragments (see text).}
\label{F5}
\end{figure}

\begin{figure}[h] 
\includegraphics[bb=19 291 559 769
,width=0.80\linewidth,clip]{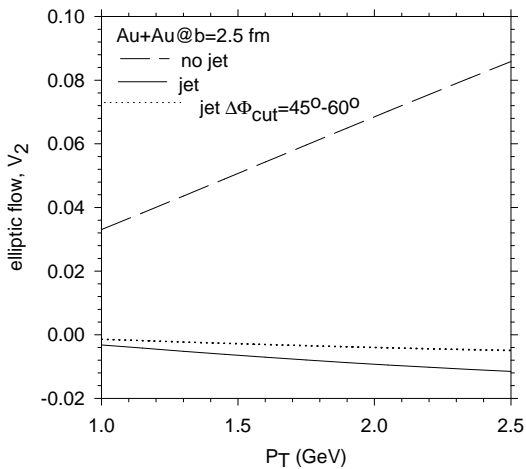}
\caption{same as in Fig.\ref{F5}, but for elliptic flow.}
\label{F6}
\end{figure}

In Fig.\ref{F5} and \ref{F6}, we have shown the simulation
results for the transverse momentum dependence of the
directed flow and the elliptic flow.
For each jet trajectories ($\phi_{prod}$,$\phi_{jet}$), we calculate
the
directed and the elliptic flow as,
 
 \begin{eqnarray} \label{eq10}
v_1 (p_T,\phi_{prod},\phi_{jet})=&&\frac{ \int^\pi_{-\pi} d\phi \frac{dN}{d^2p_T d\phi} cos(\phi - \phi_{jet}) }
{ \int^\pi_{-\pi} d\phi \frac{dN}{d^2p_T d\phi}  }\\
v_2 (p_T,\phi_{prod},\phi_{jet})=&&\frac{ \int^\pi_{-\pi} d\phi \frac{dN}{d^2p_T d\phi} cos(2(\phi - \phi_{jet})) }
{ \int^\pi_{-\pi} d\phi \frac{dN}{d^2p_T d\phi}  },
\end{eqnarray}
 
\noindent and average over $\phi_{prod}$ and $\phi{jet}$.
In Fig.\ref{F5}, $p_T$ dependence of
the directed flow $v_1$ is shown. As expected, without any quenching jet, directed flow is exactly zero (the dashed line).  
In Fig.\ref{F5},
the solid line shows the $p_T$ dependence of directed flow with a quenching jet.
With quenching jet the directed flow is non-zero and negative.  
For jet energy loss, consistent
with high $p_T$ suppression at RHIC, $\el$=14 GeV/fm, the quenching jet induces small negative directed flow $-v_1 \sim 1.5-2.5\%$  in $p_T$ range 
$p_T$= 1-2.5 GeV.  The result is not unexpected.  As discussed earlier (Fig.\ref{F4}), azimuthal distribution of $p_T$ integrated pions show enhanced production at $\phi=\pi$ and depleted production at $\phi=0$. Since directed flow is average of $cos\phi$, negative $v_1$ is expected. The results
is interesting.
Symmetry considerations require that at mid-rapidity, $p_T$ integrated $v_1$ is exactly zero. However, with a quenching jet, as seen in Fig.\ref{F5}, apparently $p_T$ integrated $v_1$ is non-zero and negative. The inconsistency can be resolved if we note that the present model neglect the trigger jet.  
Thus there is a net momentum imparted in the direction
of the jet. In a real world, the trigger jet, moving in the opposite direction,
will balance the momentum imparted by the quenching jet.
The escaped jet will
fragment in vacuum and produce particles.   
For the present discussion, particles can be classified in to
two classes, (i) fluid particles and (ii) trigger jet fragments.
Fig.\ref{F3}, depicts the directed flow from fluid particles only. 
Trigger jet fragments will produce positive $v_1$  which,   will balance, on the average, the negative $v_1$ from fluid particles.
Experimentally, as trigger jet fragments cannot be distinguished from fluid particles, on the average, one will observe net zero directed flow.
To include the effect of trigger jet fragments we correct the spectra as follows: trigger jet fragments will populate a cone  around $\phi=\phi_{jet}$. They will be balanced by fluid particles
in a cone around $\phi=\pi+\phi_{jet}$. To include the effect of trigger jet fragments, in the calculated spectra $dN/d^2p_Td\phi$, we cut out angular region
$\pm \Delta \phi_{cut}$ around $\phi=\phi_{jet}$ and $\phi=\pi+\phi_{jet}$,
and replace the cut by the angular averaged production. This
replacement will approximately include the trigger jet fragments
contribution to directed flow. This ad-hoc procedure is plagued
by the uncertainty in the value of $\Delta\phi_{cut}$. If it is small,
trigger jet fragments will not be fully accounted. If large, genuine
fluid particles will be lost.  To show the dependence of $\Delta \phi_{cut}$ on directed flow, in Fig.\ref{F5} we have shown the   directed flow from the "corrected" spectra for different values of the cut angle,  
$\Delta\phi_{cut}$=$45^\circ$,$50^\circ$,$55^\circ$, and $60^\circ$
(the dotted lines from bottom to top). 
The ad-hoc procedure to correct for the  trigger jet fragments
drastically reduces the directed flow.  For $\Delta\phi_{cut}=45^\circ$, flow is reduced by a factor of 5 or more.
For $\Delta\phi_{cut}=55^\circ-60^\circ$, , model predictions are consistent with near zero directed flow.   
The angle can be compared with the width of a jet. Recently
PHENIX collaboration studied the centrality dependence of 
the width of
the near and the away side jet \cite{Adler:2005ee}. In Au+Au collisions
the near side jet (rms) width is small,
approximately $\sigma \sim$ 0.3 rad. It is rather large for the away side jet, $\sigma \sim$ 1.1 rad. The cut angle 
$\Delta\phi_{cut} \sim 55^\circ-60^\circ$,  is approximately of the width of the away side jet.

Let us now investigate the elliptic flow. We have argued that
in presence of a jet, elliptic flow reduces. In Fig.\ref{F6}, the dashed line show the  $p_T$ dependence of elliptic flow in
evolution without any quenching jet. In b=2.5 fm Au+Au collision, 
initial asymmetry of the reaction zone produces positive elliptic flow $\sim$ 4-8\% in $p_T$ range 1-2.5 GeV. In Fig.\ref{F6}, the solid line is the elliptic flow with a quenching jet. With a quenching jet, elliptic flow is reduced
and become small negative, less than -1\%. The reason is understood. The quenching jet induces negative flow, which overrides the positive flow due to initial asymmetry and makes the overall flow negative. However,
 as in directed flow, in elliptic flow also, corrections for trigger jet fragments may be important. Indeed, it is unlikely that trigger jet fragments have an ideal dipole distribution $dN/d\phi \propto 1+cos(\phi)$ and
only neutralize the negative $v_1$ from fluid particles. Trigger jet fragments will also contribute to elliptic flow. Experiments at RHIC do indicate that even at high momentum, particles do show
azimuthal correlation like elliptic flow \cite{Adler:2002ct,Adams:2004wz}.   
In Fig.\ref{F6}, the dotted lines are the elliptic flow after correcting the spectra as indicated earlier. $v_2$ for different $\Delta\phi_{cut}$=$45^\circ$-$60^\circ$ cannot be distinguished.
With
corrections for trigger jet fragments  included  the elliptic flow become less negative and   it  reduces to near zero value. 
We conclude that the jet quenching approximately neutralises the elliptic flow in a 
0-5\% centrality Au+Au collision.
This is the main result of the present analysis. In a finite
impact parameter collision, 
in presence of a quenching jet, elliptic flow is strongly reduced.  
The result can be used to identify a jet event. As it is well known,  identifying a jet event in heavy ion collisions is problematic. Unlike in $e^+e^-$ or in $pp$ collisions, in heavy ion collisions, huge background makes it near impossible to identify a jet event. Strong reduction in elliptic flow, in presence of a jet suggests a simple and practical  way to identify a jet event. Measure the elliptic flow on a event-by-event basis. A jet event will have considerably less elliptic flow than the average. 
\begin{figure}[h] 
\includegraphics[bb=22 291 530 769
,width=0.80\linewidth,clip]{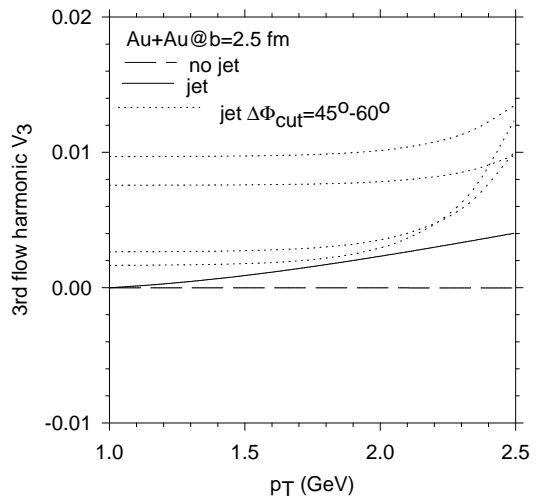}
\caption{same as in Fig.\ref{F5}, but for 3rd flow harmonic $v_3$.} \label{F7}
\end{figure}
We have also investigated the $p_T$ dependence of 3rd flow harmonic $v_3$. As mentioned in the beginning, if due to Mach 
shock wave, particle production is inhibited in $\pm 67^\circ$,
one expect some $v_3$. In Fig.\ref{F7}, our simulation results are shown. Without any quenching jet $v_3$ is identically zero. With
a quenching jet, small positive $v_3$ (less than 0.2\%) develops. $v_3$ increases if we correct the spectra as indicated above. Interestingly, unlike the elliptic flow, 3rd flow harmonic is very sensitive to the cut angle $\Delta \phi_{cut}$ and increase as
 $\Delta \phi_{cut}$ increases. For $\Delta\phi_{cut}\approx 55^\circ-60^\circ$, which approximately neutralises the directed flow (see Fig.\ref{F5}), 3rd flow harmonic $v_3\sim 1\%$. The simulation suggest that quenching jet induces small positive $v_3$ which can also be used to identify a jet event. However, detecting small $v_3\sim 0.1\%$ may not be easy experimentally.

To conclude, we   have investigated the effect of  jet
quenching on elliptic flow. We have argued that a quenching jet defines a direction 
in otherwise symmetric reaction zone and lead to negative elliptic flow even in a central b=0 collision. Negative elliptic flow induced by a quenching jet is evident in finite impact parameter collisions also. Elliptic flow is reduced drastically. For example,  explicit simulation indicate that in 0-5\% centrality Au+Au collisions, without any quenching jet,  in the $p_T$ range 1-2.5 GeV, elliptic flow is $\sim$ 4-8\%. With a quenching jet elliptic flow reduces to near zero   or small negative value.    
Large reduction in elliptic flow in presence of a jet,  can be used to identify a jet event in heavy ion collisions. 
 

\end{document}